\documentclass[
reprint,
 amsmath,amssymb,
 aps,
prl
]{revtex4-2}

\usepackage{graphicx}
\usepackage{dcolumn}
\usepackage{bm}
\usepackage{hyperref}
\usepackage{natbib}
\usepackage{mathtools}
\usepackage{physics}
\usepackage{subcaption}
\usepackage{mathrsfs}
\usepackage{multirow}
\usepackage{booktabs}

\begin{document}

\title{Quantum Big Bounce in Wheeler-DeWitt scattering theory: Ekpyrotic and LQC-like transitions}

\author{Simone Lo Franco}
 \email{simone.lofranco@uniroma1.it}
\affiliation{%
    Department of Physics, Sapienza University of Rome, Piazzale Aldo Moro, 2, 00185 Roma, Italy
}%
\affiliation{%
    INFN Section of Rome 1, Sapienza University of Rome, Piazzale Aldo Moro, 2, 00185 Roma, Italy
}%

\author{Giovanni Montani}
\email{giovanni.montani@enea.it}
\affiliation{
    ENEA, Fusion and Nuclear Safety Department, C.R. Frascati, Via E. Fermi, 45, Frascati, 00044, Roma, Italy
}%
\affiliation{
    Department of Physics, Sapienza University of Rome, Piazzale Aldo Moro, 2, 00185 Roma, Italy
}%

\begin{abstract}
We present a rigorous description of the quantum dynamics of the closed isotropic Universe in the presence of an ekpyrotic-like scalar field, expressed in terms of relativistic quantum scatterings. Working in a covariant approach to the minisuperspace, we demonstrate the quantum equivalence between parametrizations in terms of the logarithmic scale factor and the volume variable. The analogy between the Wheeler-DeWitt equation and the Klein-Gordon equation, together with a proper definition of asymptotic states, allows us to construct a relativistic quantum-mechanical scattering framework. We describe the bouncing transitions a collapsing Universe undergoes when interacting with a Dirac delta potential, which represents the limit of an infinitely steep ekpyrotic potential. Two different transition channels can be distinguished depending on whether the relational time flow is preserved or not. We show that bouncing transitions cannot be neglected for highly delocalized states in coordinate space.
\end{abstract}
\maketitle
\section{Introduction}

It was long believed that the canonical quantization of the gravitational field, as provided by the Wheeler-DeWitt equation \cite{DeWitt:1967yk,DeWitt:1967ub,DeWitt:1967uc, Misner:1973prb,Kuchar:1980ht, isham1992rv} was unable to remove the cosmological singularity \cite{Blyth:1975is,kirillov_97,Benini:2006xu,Montani:2009pc,Montani:2014cqg}. This claim originated from the observation that the evolution of mean values obtained from localized states was essentially consistent with the Ehrenfest theorem \cite{Ehrenfest:1927swx,Sakurai:2011zz}. However, it must be emphasized that localized states are destined to spread sooner or later \cite{Giovannetti:2022qje,kirillov_93}, except for very simple cases. Thus, this argument is based on extrapolating mean-value behavior toward the Big Bang, where it is typically no longer meaningful. This point of view has been overcome and put in a new light by the cosmological implementation of Loop Quantum Gravity \cite{Thiemann:2007pyv}, mainly due to the Ashtekar School \cite{Bojowald:2001,Bojowald:2002gz,Bojowald:2005epg,Ashtekar:2006rx,Ashtekar:2006wn,Ashtekar:2006es,Ashtekar:2011ni} (for possible criticisms to these approaches see \cite{Cianfrani:2010ji,Cianfrani:2011wg, Bruno:2023aco, Bruno:2023all,Bojowald:2020wuc}). In this new scenario, the Big Bounce emerges as a natural feature of the semiclassical behavior of the isotropic Universe and Bianchi models \cite{Bojowald:2004ra,Ashtekar:2011BKL,Ashtekar:2009um, Gupt:2012vi, Singh:2012zzj}, \emph{de facto} due to the discreteness of the volume operator. This feature can be recovered in a Polymer Quantum Mechanics approach to the minisuperspace \cite{Barca:2019ane,Giovannetti:2019ewe, Giovannetti:2020nte, Barca:2021qdn, Barca:2021epy} (for an $f(R)$-gravity implementation see \cite{Limongi:2025jsh}). 
\par
The removal of the cosmological singularity is certainly a significant result. Despite the singularity removal arising from the quantum nature of space-time, we can not avoid emphasizing that the established characterization of the LQC bounce proceeds primarily through the mean-value evolution of semiclassical states, and whether a quasi-classical description holds during the bouncing phase is not yet fully assessed \cite{Ashtekar:2006rx,Ashtekar:2006wn,Montani:2018uay}. Following the seminal work of Wald \cite{Wald:1993kj,Higuchi:1994vc}, where Hilbert spaces construction from a minisuperspace quantization of the Bianchi Universes is discussed, it was argued that the Wheeler-DeWitt equation for the minisuperspace can be treated in close analogy to a Klein-Gordon one through a propagator scattering theory \cite{Giovannetti:2022qje,Giovannetti:2023psb,LoFranco:2024nss,Giovannetti:2025box}. See also the relevant study on the quantum Mixmaster model in \cite{LoFranco:2025myd}. These approaches allowed the introduction of the concept of Quantum Big Bounce, i.e., the possibility for a transition from a collapse state to one of expansion (for a scattering description of the bounce in LQC in a single scalar field frequency sector, see \cite{Kaminski:2010yz}. For other approaches to singularity removal in the WDW theory,  including Bohmian trajectory and path integral methods, see \cite{PintoNeto:2012,Pinto-Neto:2021gcl,Rajeev:2021lqk,Kiefer:2019bxk, Mukherjee:2026pym, matsui:2026}). 
\par
In the present study, we follow this Quantum Field Theory formulation for a closed isotropic Universe with ekpyrotic features \cite{Khoury:2001wf,Buchbinder:2007ad,Buchbinder:2007at,Lehners:2008vx}. These features are implemented through a scalar field potential that drives the scattering process, with the field itself serving as the physical clock. Since the detailed scattering features depend on the chosen interaction potential, we model this interaction as ``instantaneous'', via a Dirac delta potential in the time-like variable. This way, we intend to capture the most genuine properties of an ekpyrotic-like scattering, i.e., a fast transient interaction. Compared to \cite{Giovannetti:2022qje,Giovannetti:2023psb,LoFranco:2024nss}, a rigorous formulation of both scattering amplitudes and asymptotic states opens the possibility of identifying two different transition channels that can yield a Quantum Big Bounce. We first show the quantum equivalence of this KG formulation under two different scale-factor parametrizations. Using wave packets, we can semiclassically distinguish the transitions occurring for a state contracting towards the interaction region. Both bouncing and non-bouncing processes can occur along two different channels: one in which the scalar field frequency sector is preserved (as in the LQC bounce), and one reversing the relational time propagation (tied to the specific scalar field potential). We eventually compute the average outcomes of both processes and discuss their dependence on wave packet parameters. In particular, we show that the Big Bounce probability is relevant for incoming states that are considerably spread in configuration space.     

\section{Covariant approach to the minisuperspace}
We start by applying the Hamiltonian formulation of General Relativity to the closed isotropic Universe, in the presence of a homogeneous massless scalar field. When homogeneous geometries are considered, the phase space of these systems is finite-dimensional (\emph{minisuperspace}). In the case under investigation, it simply consists of the isotropic scale factor, the scalar field, and their respective conjugate momenta. The classical dynamics of this system can be computed by deriving the equations of motion from the Hamiltonian $H=N\mathcal{H}$, where $N(t)$ represents the \emph{lapse function} and $\mathcal{H}$ (\emph{super-Hamiltonian}) is the only non-trivial constraint left after requiring spatial homogeneity. The dynamics of the Universe is then analogous to that of a particle in motion on a pseudo-Riemannian space. In fact, the super-Hamiltonian constraint can be written in general as 
\begin{equation}
    \mathcal{H} = G^{ab} p_a p_b+ U(q^a)\approx 0\,.
    \label{eq:sup-ham-general}
\end{equation}
Here, $q^a,p_a$ label respectively the minisuperspace variable and their conjugate momenta, $G_{ab}$ is the minisupermetric, while $U(q^a)$ accounts for potential terms. In the case of a closed isotropic model, the three-space curvature induces a potential term $U_K$, introducing a maximal volume in the Universe's dynamics. We will later consider a contribution $U_s$ generated by the scalar field ekpyrotic potential. The Hamiltonian written in the form of Eq.~(\ref{eq:sup-ham-general}) is covariant under generic reparametrizations of the minisuperspace $q^{a'}=q^{a'}(q^a)$. Denoting the isotropic scale factor by $a(t)$ and the scalar field by $\phi$ \footnote{We adopt the rescaled scalar field $\phi \to \sqrt{4\pi/3}\,\phi$ throughout.}, the dynamics of this model is often solved in terms of either the \emph{logarithmic scale factor} $\alpha=\log{a}$ or the \emph{volume variable} $v=a^3$. At the classical level, these two parametrizations are equivalent, as they are linked through the canonical transformation $\alpha=\log(v^{1/3}),\, p_\alpha = 3vp_v$. Hamilton's equations of motion can be solved in a gauge-independent fashion by considering one of the degrees of freedom (monotonic with respect to the coordinate time $t$) to be the system's internal time. Here, we choose $\phi$ to play that role, since neither $\alpha,v$ nor $\phi$ are \emph{a priori} privileged \cite{Bamonti:2025}. Therefore, when only the curvature potential is present in $\mathcal{H}$, the evolution of the volume variable with respect to the internal time reads as
\begin{equation}
    v(\phi)=v_{max}\, \text{sech}^{3/2}\left[ 2 (\phi-\phi_0)\right]\,,
    \label{eq:v_phi_cl}
\end{equation}
where $v_{max}\equiv v(\phi_0)=[2 |p_\phi|/(3\pi)]^{3/2}$ is the maximal volume and $\phi_0$ fixes the instant at which the turning point occurs. Eq.~(\ref{eq:v_phi_cl}) coincides with $v(t)$ whenever $N(t)$ is fixed to yield $d\phi/dt = \text{sgn}(p_\phi)$. While the $\alpha$ and $v$ parametrization are equivalent at the classical level, this relation is not guaranteed at the quantum level. Therefore, we check whether the Bogolyubov transformation linking the $\alpha$ and $v$ states is unitary. Moving to the quantum level, the super-Hamiltonian constraint in Eq.~(\ref{eq:sup-ham-general}) is recast in the form of the covariant Wheeler-DeWitt equation
\begin{equation}
    \left[ -\frac{1}{\sqrt{-G}}\partial_a(\sqrt{-G}\,G^{ab}\partial_b) + \, U(q) \right] \Psi(q^a)=0\,,
    \label{eq:wdw-cov-general}
\end{equation}
where $G=\det{G_{ab}}$ and $\Psi$ is the \emph{Universe's wave function}. This equation can be equipped with the following conserved product 
\begin{equation}
    (\Psi_1,\Psi_2) = -i\int_\Sigma d\Sigma\,\sqrt{-h}\, \Psi_1^* \overleftrightarrow{\partial_a} \Psi_2\, n^a\,,
    \label{eq:inner-prod-cov-general}
\end{equation}
where $\Sigma$ denotes a space-like Cauchy hypersurface, $\sqrt{-h}$ the determinant of the minisupermetric $h_{ij}$ induced on this hypersurface, $n^a$ a time-like vector normal to $\Sigma$, and $\Psi_1^* \overleftrightarrow{\partial_a} \Psi_2=\Psi_1^* \partial_a \Psi_2 - (\partial_a\Psi_1^*)\Psi_2$. In the absence of the ekpyrotic potential, the independent frequency modes satisfying Eq.~(\ref{eq:wdw-cov-general}) in both $\alpha$ and $v$ parametrization are 
\begin{subequations}
    \begin{equation}
        f^\pm_k(\phi,\alpha)=N_k\, e^{\pm ik\phi}K_{\frac{ik}{2}}(\Lambda e^{2\alpha})\,,
        \label{eq:sol-alpha}%
    \end{equation}
    \begin{equation}
        g^\pm_\omega(\phi,v)=N_\omega\, e^{\pm i\omega\phi}K_{\frac{i\omega}{2}}(\Lambda v^{\frac{2}{3}})\,,
        \label{eq:sol-v}%
    \end{equation}
    \label{eq:sol-alpha-v}%
\end{subequations}
where $k,\omega>0$ label the ``energy'' eigenvalues, $N_\nu=[\sinh{(\pi\nu/2)}/(2\pi^2)]^{1/2}$ is a normalization constant, $\Lambda=3\pi/4$, and $K_{\nu}(z)$ denotes the modified Bessel function of the second kind (Macdonald function). These solutions satisfy the orthonormality relations $(f^\pm_{k'},f^\pm_k)=\pm\delta(k-k')\,,\,(f^\pm_{k'},f^\mp_k)=0\,;\, (g^\pm_{\omega'},g^\pm_\omega)=\pm\delta(\omega-\omega')\,,\,(g^\pm_{\omega'},g^\mp_\omega)=0$, and they obey the boundary condition of the closed Universe, i.e., $f^\pm_k,g^\pm_\omega \to 0$ as $\alpha,v \to +\infty$. Therefore, they both form a basis, and it is possible to compute the coefficients of the Bogolyubov transformation by projecting the two different modes after applying the reparametrization $v=\exp{3\alpha}$, i.e., $\mathcal{A}_{k,\omega }\equiv (g^+_\omega,f^+_k)\,,\, \mathcal{B}_{k,\omega}\equiv-(g^-_\omega,f_k^+)$. The results show that these coefficients are $\mathcal{A}_{k,\omega} =\delta(\omega-k), \mathcal{B}_{k,\omega}=0$ (see the Appendix). Therefore, the Bogolyubov transformation is trivially unitary, and the $\alpha$ and $v$ representations are physically equivalent also in a KG formulation of the WDW theory. This result is relevant for two reasons: first, we make contact with the previous two scattering studies for the isotropic Universe \cite{Giovannetti:2023psb,LoFranco:2024nss}; furthermore, such a result is interesting when compared with the LQC formalism. In fact, the discrete nature of gravitational degrees of freedom, as simply seen in a PQM approach \cite{Barca:2021qdn,Giovannetti:2020nte}, breaks this equivalence, otherwise holding in the continuum limit. From now on, throughout the work, we will focus only on the $v$ parametrization. 

\section{Universe's semiclassical states}
In Quantum Cosmology, the fundamental states of the theory must allow a meaningful localization in ``coordinate'' and ``momentum'' space. As the frequency modes in Eq.~(\ref{eq:sol-v}) do not satisfy such a requirement, we construct localized wave packets in the form
\begin{equation}
    \psi^\pm(\phi,v)=\int_\mathbb{R_+} d\omega\, A(\omega)\,g^\pm_\omega(\phi,v)\,.
    \label{eq:wp}
\end{equation}
We choose the weight function $A(\omega)$ to be
\begin{equation}
    A_{\Omega,\Phi}(\omega;\sigma) = \frac{N(\sigma)}{\sqrt{\Omega}}\exp{ -\frac{[\log(\omega/\Omega)]^2}{2\sigma^2} -i\omega\Phi}\,,
    \label{eq:lognorm-weight}
\end{equation}
where $N(\sigma)=( \sigma\sqrt{\pi}\exp{\sigma^2/4})^{-1/2}$ is a normalization factor. The reason behind this choice is mainly that wave packets built from this weight satisfy the normalization condition $(\psi^\pm_{\Omega,\Phi},\psi^\pm_{\Omega,\Phi})=\pm1$, and they form an over-complete set of states in the parameter space with measure $\exp{-\sigma^2/4}/(2\pi)\,d\Omega d\Phi$ \footnote{It can be checked that $\exp{-\sigma^2/4}/(2\pi)\int d\Omega d\Phi\ A^*_{\Omega,\Phi}(\omega';\sigma)A_{\Omega,\Phi}(\omega;\sigma)=\delta(\omega'-\omega)$.}. Therefore, they can be easily implemented in a scattering description. We can now introduce the probability densities for the positive and negative frequency subsets as
\begin{equation}
    \mathcal{P_\pm}(\phi,v) = \mp \frac{i}{3} \psi^{\pm\, *} (\phi,v)\, \overleftrightarrow{\partial_\phi}\, \psi^\pm (\phi,v)\,,
    \label{eq:p-densities}
\end{equation}
which gives the conditional probability of finding the Universe with some volume $v$ at a given time $\phi$. Despite the well-known problems with the probabilistic interpretation of KG-like inner products, as long as $\sigma$ is sufficiently small, we can treat the densities (\ref{eq:p-densities}) as positive definite \cite{Giovannetti:2022qje}. Regarding the existence of two frequency-split subsets, we stick with Feynman's interpretation that positive frequency states propagate forward in time $\phi$ while negative frequency states are backward propagating, as first investigated in \cite{LoFranco:2024nss}. This interpretation is well-grounded in the minisuperspace, since the classical evolution itself can be described forward or backward in the relational time $\phi$, depending on the choice of $p_\phi$. The wave packet parameters are related to physical quantities in the following way: $\Omega$ and $\sigma$ control the mean value of the energy operator $\hat{p}_\phi=-i\partial_\phi$, i.e., $\overline{p}_\phi=\Omega \exp{3\sigma^2/4}$ (note that $\overline{p}_\phi \approx \Omega$ for $\sigma \ll 1$), while $\sigma$ alone fixes the relative uncertainty $\delta p_\phi/\overline{p}_\phi = \sqrt{\exp{\sigma^2/2}-1}$. Instead, $\Phi$ determines the instant at which the closed-Universe turning point occurs semiclassically. Therefore, $\Omega,\Phi$ are the semiclassical analogs of $p_\phi,\phi_0$ from Eq.~(\ref{eq:v_phi_cl}). Note that the quantum parameter $\Phi$ acts differently depending on the frequency's sign, a consequence of the different direction of the internal time evolution. 

\section{Scattering with the interaction potential}
We now move to describe how the quantum closed Universe is affected by the emergence of an ekpyrotic-like self-interaction potential $U_s(\phi)$. The idea is that such a potential is relevant for a brief time near the singularity, and is adiabatically switched off at early/late times. In the asymptotically free regions, the Universe is described by the wave functions in Eq.~(\ref{eq:wp}). When $U_s(\phi)$ becomes non-negligible, the Universe undergoes a purely quantum phase where it is described by a positive and negative frequency superposition. After some time, the Universe will perform a transition to another meaningful semiclassical state. Transition amplitudes can be evaluated through the propagator scattering theory of Relativistic Quantum Mechanics \cite{Bjorken:1965sts}. Here, however, we will approximate the ekpyrotic potential to compute non-perturbative transition amplitudes without directly implementing it. We consider the scalar field potential in the form $U_s(\phi)=-6\eta\pi^3\, \delta(\phi)$, where $\eta$ is a coupling parameter. Such a choice captures the features of an infinitely steep ekpyrotic potential, and it can be recovered from complete expressions in appropriate shape parameter limits.  In this setting, the solutions of the full WDW equation are everywhere in the form of Eq.~(\ref{eq:sol-v}) except for $\phi=0$. We want to describe the transitions that an incoming positive-frequency $g^+_\omega$ state undergoes when interacting with the scattering potential. To do so, we split the solutions of the full equations into two branches: one for $\phi<0$ and the other for $\phi>0$. In the former branch, given the boundary conditions of the Feynman propagator, both the incoming state $g^+_\omega$ and the reflected backward-propagating negative-frequency component are present. In the $\phi>0$ branch, we find only the transmitted positive-frequency components (negative-frequency states in this region would propagate freely from $\phi \to +\infty$, corresponding to a many-Universes interaction). Therefore, we consider solutions of the complete WDW equation $\Psi$ to be in the form
\begin{equation}
    \Psi_{<} = g^+_\omega + \int_{\mathbb{R}_+} d\omega'\, \mathcal{R}_{\omega',\omega}\, g_{\omega'}^-\,,\quad \Psi_{>} = \int_{\mathbb{R}_+} d\omega'\, \mathcal{T}_{\omega',\omega}\, g^+_{\omega'}\,,
    \label{eq:wdw-full-sol}
\end{equation}
where subscripts $<,>$ are shorthands for $\phi<0,\phi>0$. $\mathcal{R}$ and $\mathcal{T}$ correspond to the probability amplitude of $g^+_\omega \to g^-_{\omega'}$ and $g^+_\omega \to g^+_{\omega'}$ transitions respectively. In order for Eq.~(\ref{eq:wdw-full-sol}) to satisfy the full WDW equation, we must impose both continuity and a jump condition for the first derivative w.r.t. $\phi$ in $\phi=0$. From these two equations, we find the following expression for $\mathcal{R}$ and $\mathcal{T}$
\begin{subequations}
    \begin{equation}
        \mathcal{R}_{\omega',\omega} = iU_{\omega',\omega} + i\int_{\mathbb{R}_+} d\omega''\, U_{\omega',\omega''} \mathcal{R}_{\omega'',\omega}\,,
        \label{eq:R}%
    \end{equation}
    \begin{equation}
        \mathcal{T}_{\omega',\omega} = \delta(\omega'-\omega) + \mathcal{R}_{\omega',\omega}\,,
        \label{eq:T}%
    \end{equation}
    \label{eq:R-T}%
\end{subequations}
Where $U_{\omega',\omega}=2\eta\pi^3 \int_{\mathbb{R}_+} dv\, v\, \chi_{\omega'}\chi_{\omega}\,,\ \chi_\omega=N_\omega K_{i\omega/2}(\Lambda v^{2/3})$ (see the Appendix). $\mathcal{R}$ coefficients satisfy a Fredholm integral equation, while $\mathcal{T}$ coefficients are composed of the no-interaction Dirac delta term plus an interaction term that is exactly $\mathcal{R}$. Hence, the interaction parts of two channels coincide. KG current conservation shows that $\mathcal{R}$ and $\mathcal{T}$ are the coefficients of a unitary operator, and they can be interpreted as non-perturbative differential probability amplitudes. From numerical computations, $|\mathcal{R}_{\omega',\omega}|^2$ initially grows with $\omega$ and stabilizes asymptotically. For $\omega \gg 1 $, the most probable outcome is an elastic transition $\omega' = \omega$. As we developed a way to compute non-perturbative transition amplitudes for the independent solutions of the WDW equation, we are left with the task of interpreting these transitions in the cosmological setting. Given the presence of the curvature potential, $\hat{D}_v$ eigenstates are mixed in the solutions (\ref{eq:sol-v}), and therefore we cannot distinguish semiclassically between expansion and contraction states. However, despite this additional technical difficulty, the introduction of the curvature potential naturally bounds the volume contribution from the scattering potential, ensuring that $U_{\omega',\omega}$ is a finite quantity.

\begin{figure*}[t]
    \centering
    \includegraphics[width=1\linewidth]{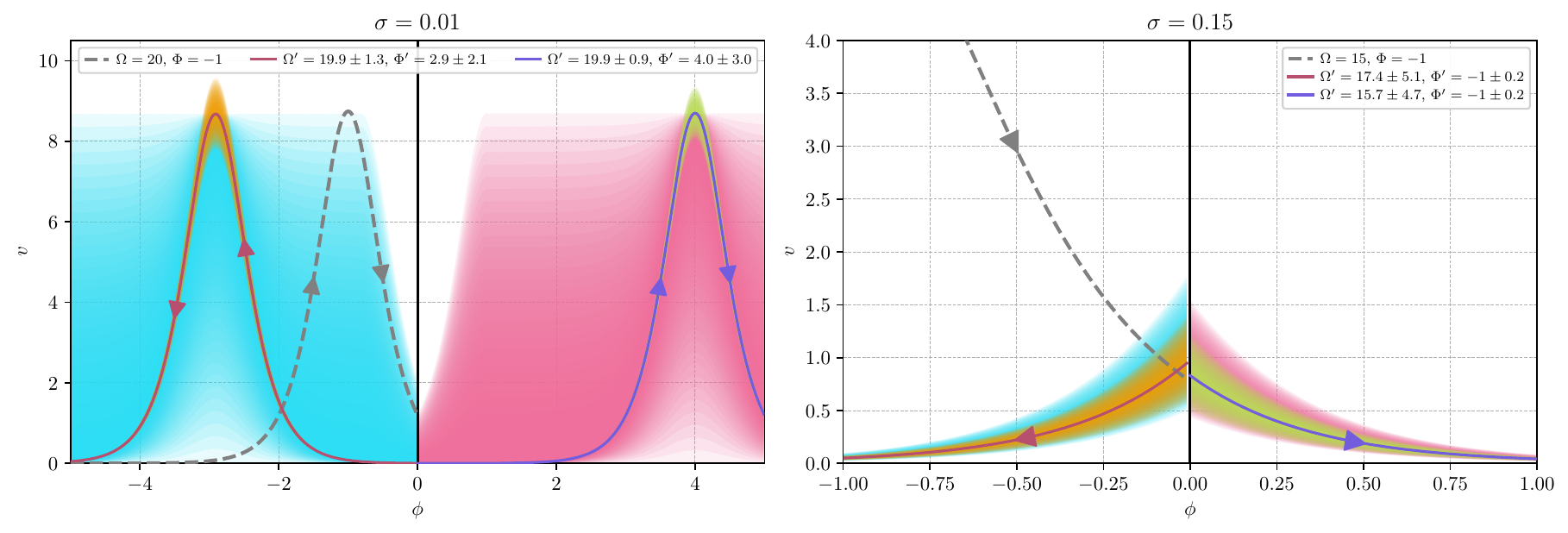}
    \caption{Classical trajectories computed for in (dashed line) and out (solid lines) states, representing a bouncing configuration (left) and an ekpyrotic contraction (right). Positive and negative frequency states are distinguished by the arrow of their internal time evolution. We plot the regions in which all the trajectories fall in the interval $[\langle \Omega'\rangle -\Delta\Omega', \langle \Omega'\rangle + \Delta\Omega']$ at $\langle \Phi' \rangle$ fixed (gold and green), and $[\langle \Phi'\rangle -\Delta\Phi', \langle \Phi'\rangle + \Delta\Phi']$ at $\langle \Omega' \rangle$ fixed (cyan and pink).}
    \label{fig:1}%
\end{figure*}
    
\section{The Big Bounce transition}
We now characterize the transitions of an incoming localized state $\psi^+_{\Omega,\Phi}$ driven by the scalar field potential, and compute the related amplitudes. As $U_s$ is relevant around $\phi=0$, any negative/positive frequency state with $\Phi < 0$ can be considered to approach the scattering region while contracting (vice versa for $\Phi > 0$). Therefore, a transition from a state $\Phi < 0$ to another $\Phi' > 0$ represents a Quantum Big Bounce indeed. When this transition occurs in the same frequency subset, the arrow of time remains fixed during the process. This is the case of the LQC Big Bounce, where discrete gravitational degrees of freedom alone drive singularity avoidance without internal time reversal. Hence, we refer to the $+,\Omega,\Phi < 0 \to +,\Omega',\Phi'> 0$ transition as the \emph{Frequency Sign Preserving} (FSP) \emph{Bounce}. The $+,\Omega,\Phi<0 \to -,\Omega',\Phi'> 0$ transition is instead called the \emph{Frequency Sign Reversing} (FSR) \emph{Bounce}, where the internal time flow is reversed. Notice that such a process is strictly related to the potential under consideration. The Dirac delta form yields two transition terms for FSP and FSR processes that are equal. If we instead consider a finite interaction time, the transition terms for the two processes would be modulated accordingly. Regarding $+,\Omega,\Phi < 0 \to \pm,\Omega',\Phi'< 0$ transitions, by the same reasoning, they represent transitions after which the Universe keeps contracting towards the singularity. The reflection $\mathcal{R}_\sigma (\Omega',\Phi';\Omega,\Phi)$ and transmission coefficients $\mathcal{T}_\sigma (\Omega',\Phi';\Omega,\Phi)$ are obtained by integrating Eqs. (\ref{eq:R-T}) together with the incoming and outgoing wave packets (\ref{eq:lognorm-weight}), and they are well-normalized probability amplitudes. We consider $\sigma$ to be the same for in and out states. From $|\mathcal{R}|^2$ and $|\mathcal{T}|^2$, we can extract some relevant quantities: the probabilities of FSP/FSR bounce/contraction $\mathcal{P}^{(\mathcal{T/R})}_{B/C}$ and the outgoing state mean values and variances related to these processes $\langle \Omega'\rangle\,,\,\langle \Phi'\rangle\,,\Delta\Omega'\,, \Delta\Phi'$ (see the Appendix). We restrict our attention to semiclassical incoming states, i.e., those whose semiclassical $\overline{v}_{max} \propto \Omega^{3/2}$ is far from the interaction volume (hence, $\Omega \gg 1$). Some representative results are presented in Table \ref{tab-1}. The partition of bounce and contraction probability is mainly controlled by wave packet spread. In particular, states that are well localized in the $v$ space ($\sigma$ larger) have a smaller (even negligible) probability of bouncing off the potential. On the other hand, for narrow wave packets in $\omega$-space ($\sigma$ smaller), the occurrence of a Big Bounce is equally likely to keep contracting. Regarding the $\Phi$-dependence, incoming states whose maximal turning point instant in $\phi$ is placed far from the interaction have a smaller bounce probability. Given this dependence on $\sigma$, states that keep collapsing after the interaction have small fluctuations on $\Phi'$ but a greater uncertainty on $\Omega'$. On the contrary, bouncing distributions are broadly spread in $\Phi'$-space, but peaked in $\Omega'$-space. Despite reducing the potential to a Dirac delta function, notice that we still find signatures of an ekpyrotic phenomenology for $v$-localized states as a higher $\langle\Omega'\rangle$, on average, compared to the incoming $\Omega$. It is characteristic of negative potentials to increase the $\phi$ kinetic term during the contraction. Instead, we find $\langle \Omega' \rangle \lesssim \Omega$ for bouncing transitions. In Fig.~\ref{fig:1}, we provide a graphical representation of both the bounce and the ekpyrotic contraction by plotting semiclassical trajectories for incoming and outgoing states. We consider a case in which the bounce probability is relevant, and one in which the contraction is the dominant outcome. 

\begin{table}[htbp]
    \centering
    \footnotesize 
    \setlength{\tabcolsep}{2pt} 
    \begin{tabular}{@{}cc c c c c c@{}}
        \toprule
        ($\Omega,\Phi,\sigma$) & \textbf{Type} & $\mathcal{P}_B$ & $\langle \Omega'\rangle_{B}$ & $\langle \Phi'\rangle_{B}$ & $\langle \Omega'\rangle_{C}$ & $\langle \Phi'\rangle_{C}$ \\
        \midrule
        \multirow{2}{*}{$20,-1,0.01$} 
        & FSP & $0.27$ & $19.9(9)$ & $4.1(30)$ & $20.1(9)$ & $-4.8(32)$ \\
        & FSR & $0.15$ & $19.9(13)$ & $2.9(22)$ & $20.2(15)$ & $-2.9(22)$ \\
        \midrule
        \multirow{2}{*}{$20,-2,0.01$} 
        & FSP & $0.21$  & $19.9(7)$ & $3.7(29)$ & $20.1(8)$ & $-5.3(35)$ \\
        & FSR & $0.13$  & $19.9(13)$ & $2.9(22)$ & $20.3(15)$ & $-2.9(22)$ \\
        \midrule
        \multirow{2}{*}{$20,-1,0.15$} 
        & FSP & $10^{-7}$ & $17.7(34)$ & $0.1(1)$ & $21.4(39)$ & $-1.4(3)$ \\
        & FSR & $10^{-3}$ & $18.6(27)$ & $0.1(1)$ & $21.2(33)$ & $-1.0(3)$ \\
        \midrule
        \multirow{2}{*}{$15,-1,0.03$} 
        & FSP & $0.09$    & $14.7(17)$ & $1.9(15)$ & $15.4(14)$ & $-2.7(16)$ \\
        & FSR & $0.23$    & $14.9(12)$ & $1.3(10)$ & $15.5(16)$ & $-1.4(10)$ \\
        \midrule
        \multirow{2}{*}{$15,-2,0.03$} 
        & FSP & $0.05$ & $15.1(16)$ & $1.2(12)$ & $15.3(14)$ & $-3.1(17)$ \\
        & FSR & $0.12$ & $15.1(12)$ & $1.2(10)$ & $15.7(17)$ & $-1.5(11)$ \\
        \midrule
        \multirow{2}{*}{$15,-1,0.15$} 
        & FSP & $10^{-4}$ & $14.3(23)$ & $0.2(2)$ & $16.5(27)$ & $-1.7(5)$ \\
        & FSR & $0.06$ & $14.9(15)$ & $0.3(3)$ & $16.0(22)$ & $-0.9(5)$ \\
        \bottomrule
    \end{tabular}
    \caption{Results extracted from wave packet transition amplitudes for different choices of initial state parameters. Numbers in parentheses denote the uncertainties $\Delta \Omega',\Delta\Phi'$ in the last reported digits. Here $\eta=2$.
    }
    \label{tab-1}
\end{table}

\section{Conclusions}
By implementing a covariant approach to the minisuperspace, we investigated the dynamics of a quantum closed Universe in the presence of a scalar field subjected to an ekpyrotic-like potential. Following previous proposals \cite{LoFranco:2024nss, LoFranco:2025myd}, the frequency separation of the solutions of the WDW equation distinguishes the internal time direction that is followed during the semiclassical evolution. By invoking a quantum relativistic scattering theory, we described transitions of a contracting Universe, arising when an ekpyrotic potential for the scalar field is relevant in an infinitely small time frame. We identified two different bouncing scenarios. In one case, the transition occurs along a fixed direction of the internal time arrow, as it happens in LQC. The Big Bounce can also occur thanks to a reversal of the internal time arrow, the Ekpyrotic Bounce, and it is peculiar to scalar field potentials. The results showed that bouncing transitions have significant probabilities whenever the Universe approaches the potential while highly delocalized in the volume space. States' localization also dictates how the outgoing states are distributed. These range from bouncing scenarios, where energy is lost on average, to contracting ones, where the Universe keeps collapsing with higher mean energy, as is characteristic of negative potentials. Let us stress the fact that here $\sigma$ is not a free parameter to fix, but determined by the preceding dynamics. Ekpyrotic potentials are considered in anisotropic models to drive an isotropization mechanism. When considering an isotropic model, we are effectively treating anisotropies as already washed out, but information about their effects can be inferred from $ \sigma$. In fact, anisotropic models are known to spread in coordinate space due to a non-linear dispersion relation. Therefore, smaller $\sigma$ regimes correspond to the case where anisotropies had already dominated the evolution before approaching the ekpyrotic potential. On the other hand, for high $\sigma$, anisotropies are controlled, and the Universe keeps contracting after an ekpyrotic energy bump. The major merit of the present study is demonstrating how a bouncing cosmology can emerge already at the level of the metric approach, based on the analogy with relativistic scattering, without the need for additional ad hoc singularity-avoidance boundary conditions. 
Our result has a more general conceptual value than previous results, obtained on a classical level via specific initial conditions \cite{Barrow:1980,Page:1984qt,Cornish:1998,Gordon:2003} or extending the geometrical picture to the presence of torsion \cite{Bombacigno:2021bpk}, or on a quantum level, by introducing particular source terms \cite{Alvarenga:2001nm,Batista:2001ti}. Furthermore, it is worth noting that our quantum bounce is more complementary than competitive to the results of the Ashtekar School \cite{Ashtekar:2011ni,Ashtekar:2006es}, because the scattering paradigm we adopt here could also be exported to the loop quantum sector, in those cases when the semi-classical picture has to be abandoned in favor of a full quantum scenario. In Loop Quantum Cosmology, the bounce is certainly a full quantum process, as implied by the geometrical operator discrete spectrum \cite{Rovelli:1994ge}, but its description, in the absence of a quasi-classical limit, requires a similar approach to that one we are proposing here.

\acknowledgments{S.L.F. would like to thank the TAsP INFN initiative (Rome 1 section) for support.}

\appendix
\section{\label{app:bglbv}The Bogolyubov transformation}
In this section, we compute the Bogolyubov coefficients for the solutions in Eq.~(\ref{eq:sol-alpha-v}). We start by writing the super-Hamiltonian for both $\alpha$ and $v$ parametrizations, i.e.,
\begin{subequations}
    \begin{equation}
        \mathcal{H}_\alpha = \frac{e^{-3\alpha}}{3\pi}\left[  p_\phi^2-p_\alpha^2 -e^{3\alpha}U_K(\alpha) + e^{6\alpha}U_s(\phi) \right],
        \label{eq-app:sup-ham-alpha}%
    \end{equation}  
    \begin{equation}
        \mathcal{H}_v=\frac{v^{-1}}{3\pi}\left[p_\phi^2-9v^2p_v^2- v\,U_K(v) + v^2\,U_s(\phi) \right].
    \label{eq-app:sup-ham-v}%
\end{equation}
\end{subequations}
Here, $U_K$ denotes the spatial curvature potential, while $U_s$ arises from the ekpyrotic character of the scalar field. From these equations, we can infer that the respective minisupermetrics are
\begin{subequations}
    \begin{equation}
        G^{ab} = \frac{e^{-3\alpha}}{3\pi} \text{diag}\left[ 1,-1 \right]\,,
    \end{equation}
    \begin{equation}
        G^{a'b'} = \frac{v^{-1}}{3\pi} \text{diag}\left[ 1,-9v^2 \right]\,,
    \end{equation}
    \label{eq-app:minisupermetrics}%
\end{subequations}

with $a,b=\{\phi,\alpha\}\,,\,a',b'=\{\phi,v\}$. In the regions where $U_s$ can be neglected, the WDW equation (\ref{eq:wdw-cov-general}) can be written as
\begin{subequations}
    \begin{equation}
            \left[-\partial_\phi^2 + \partial_\alpha^2 - 4\Lambda^2 e^{4\alpha}\right] \psi(\phi,\alpha)=0\,,
        \label{eq-app:wdw-alpha}%
    \end{equation}
    \begin{equation}
        \left[-\partial_\phi^2 - 9\hat{D}^2_v - 4\Lambda^2 v^{4/3}\right] \psi(\phi,v)=0\,,
        \label{eq-app:wdw-v}%
    \end{equation}
    \label{eq-app:wdw-alpha-v}%
\end{subequations}
where $\hat{D}_v = -iv\partial_v$ and $\Lambda=3\pi/4$. We can immediately notice that the WDW equation (\ref{eq-app:wdw-alpha}) in the $\alpha$ representation is perfectly analogous to a Klein-Gordon equation. In fact, in this case, the minisupermetric is conformal to a 1+1-dimensional Minkowski metric. Regarding the $v$ parametrization, instead, the kinetic term is more involved, due to the non-linear nature of the reparametrization. To treat $\phi$ as the internal time at the quantum level, we define the products (\ref{eq:inner-prod-cov-general}) over the hypersurfaces at $\phi=const$, so that $n^a=(\exp{-3\alpha/2}/\sqrt{3\pi},0)\,,\, \sqrt{-h_\alpha} = \sqrt{3\pi}\exp{3\alpha/2}$ and $n^{a'}=((3\pi v)^{-1/2},0)\,,\, \sqrt{-h_v} = (27\pi v)^{-1/2}$. Explicitly we have
\begin{subequations}
    \begin{equation}
        (\psi_1,\psi_2)_\alpha = -i \int_\mathbb{R} d\alpha\ \psi_1^* \overleftrightarrow{\partial_\phi} \psi_2\,,
    \label{eq-app:prod-alpha}%
    \end{equation}
    \begin{equation}
        (\psi_1,\psi_2)_v = -\frac{i}{3} \int_\mathbb{R_+} \frac{dv}{v}\ \psi_1^* \overleftrightarrow{\partial_\phi} \psi_2\,.
    \label{eq-app:prod-v}%
    \end{equation}
    \label{eq-app:prod}%
\end{subequations}
We define the Bogolyubov transformation, linking the solutions (\ref{eq:sol-alpha-v})
    \begin{equation}
        f^+_k(\phi,\alpha) = \int_{\mathbb{R}_+} d\omega \left[ \mathcal{A}_{k,\omega } \,g_\omega^+ + \mathcal{B}_{k,\omega} \,g_\omega^-\right]\,,
        \label{eq-app:bogo_f_plus}%
    \end{equation}
    \label{eq-app:bogo}%
where $\mathcal{A}_{k,\omega }\equiv (g^+_\omega,f^+_k)_\alpha\,,\, \mathcal{B}_{k,\omega}\equiv-(g^-_\omega,f_k^+)_\alpha$. The transformation rules for $f^-_k$ and $g^-_\omega$ can be found after noting that $f^-_k=(f^+_k)^*\,,\,g^-_\omega=(g^+_\omega)^*$. Using the orthogonality relation of the Macdonald functions $2\nu\sinh{\pi\nu}\int_\mathbb{R_+} \frac{dx}{x} K_{i\nu'}(x)K_{i\nu}(x)=\pi^2\delta(\nu'-\nu)$ \cite{PASSIAN2009380} we find the result shown in the main text $\mathcal{A}_{k,\omega} =\delta(\omega-k), \mathcal{B}_{k,\omega}=0$.

\section{\label{app:Sm-basis} S-matrix coefficients for independent solutions}
We show how to derive the reflection and transmission coefficients introduced in Eq.~(\ref{eq:wdw-full-sol}) from the full WDW equation 
\begin{equation}
    \left[-\partial_\phi^2 - 9\hat{D}^2_v - 4\Lambda^2 v^{4/3} - 6\eta\pi^3 v^2 \delta(\phi)\right] \Psi=0\,.
    \label{eq-app:full-WDW-dirac}
\end{equation}
Given the form of the interaction potential, we must impose the continuity of $\Psi$ at $\phi=0$ and a jump condition on $\partial_\phi \Psi|_{0}$ that can be obtained by integrating Eq.~(\ref{eq-app:full-WDW-dirac}) over an infinitesimal interval around $\phi=0$. In terms of solutions (\ref{eq:wdw-full-sol})
\begin{subequations}
    \begin{equation}
        \Psi_{<}|_{0_-}=\Psi_{>}|_{0_+}\,,
    \end{equation}
    \begin{equation}
        \partial_\phi\Psi_{<}|_{0_-}-\partial_\phi \Psi_{>}|_{0_+}  = 6\eta\pi^3 v^2 \Psi|_0\,.
    \end{equation}
    \label{eq-app:cont_jump}
\end{subequations}

\begin{figure}
    \centering
    \includegraphics[width=1\linewidth]{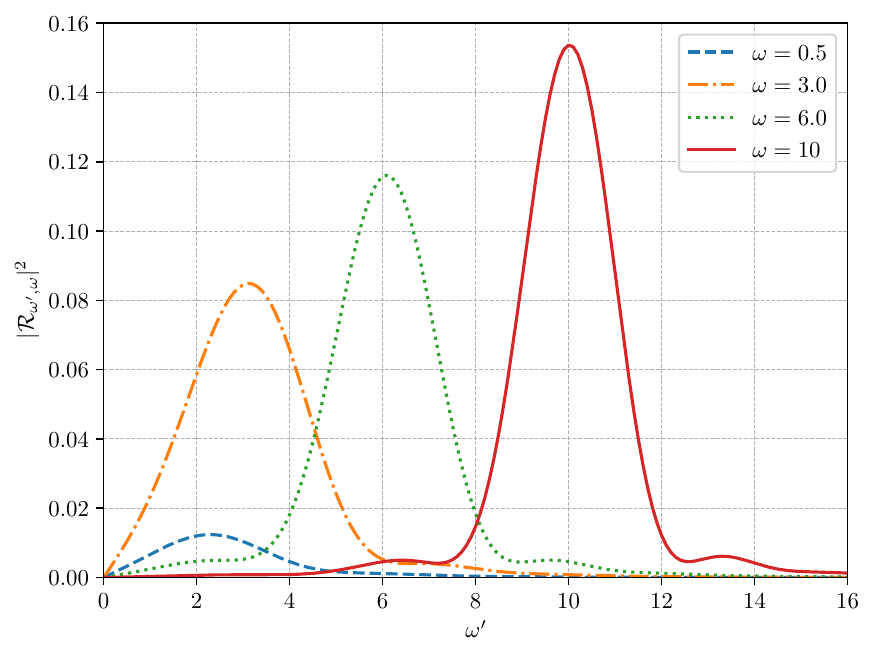}
    \caption{Plot of $|\mathcal{R}_{\omega',\omega}|^2$ as a function of outgoing $\omega'$ for different choices of the incoming $\omega$. We considered $\eta=2$.}
    \label{fig2}
\end{figure}

By substituting the explicit form of Eq.~(\ref{eq:wdw-full-sol}) and projecting over the orthonormal volume kernels $\chi_{\omega}$, we obtain Eqs. (\ref{eq:R-T}) where $U_{\omega',\omega} \equiv 2\eta\pi^3\int_{\mathbb{R}_+} dv\,v\, \chi_{\omega'}\chi_{\omega}$ has the known closed form (Eq.~6.576 4. from \cite{Gradshteyn})
\begin{equation}
    \begin{split}
        U_{\omega',\omega} =& \frac{2\eta}{9} [4+(\omega'-\omega)^2][4+(\omega'+\omega)^2]\times \\
        &\times\frac{\sqrt{\sinh{(\pi\omega'/2)} \sinh{(\pi\omega/2)}}}{\cosh{(\pi\omega'/2)}+\cosh{(\pi \omega/2)}}\,.
    \end{split}
\end{equation}
Notice that despite $\partial_\phi\Psi|_0$ being discontinuous, the KG norm is still conserved across the interaction since $U_s(\phi)$ is a real potential. Moreover, KG current conservation $(\Psi_< ,\Psi_<)=(\Psi_> ,\Psi_>)$ yields the relations
\begin{equation}
    \int_{\mathbb{R}_+} d\omega''\, (\mathcal{T}_{\omega'',\omega'}^* \mathcal{T}_{\omega'',\omega} + \mathcal{R}_{\omega'',\omega'}^* \mathcal{R}_{\omega'',\omega}) = \delta(\omega'-\omega)\,.
\end{equation}
Hence $\mathcal{T}_{\omega',\omega}\ \text{and}\ \mathcal{R}_{\omega',\omega}$ are the coefficients of a unitary scattering operator. Through Eq.~(\ref{eq:T}), this condition can be rewritten in terms of $\mathcal{R}$ only in the operatorial form $\hat{\mathcal{R}}+\hat{\mathcal{R}}^\dagger = -2 \hat{\mathcal{R}}^\dagger\hat{\mathcal{R}}$, which resembles the optical theorem (notice that there is an $i$ prefactor inside $\mathcal{R}$). Eq.~(\ref{eq:R}) can be solved by numerical computations, and in Fig.~(\ref{fig2}), we plot $|\mathcal{R}_{\omega',\omega}|^2$ as a function of $\omega'$ for different choices of $\omega$.

\section{\label{app:Sm-wp} S-matrix coefficients for wave packets}
When considering wave packets as initial and final states of some scattering process, the S-matrix elements associated with them can be obtained directly from those relative to the independent modes $g^\pm_\omega$ by integrating them with the incoming and outgoing wave packets, i.e.,
\begin{subequations}
    \begin{equation}
        \mathcal{T}_{\Upsilon';\Upsilon} = \int_{\mathbb{R}^2_+} d\omega'd\omega\, A^*_{\Omega',\Phi'}(\omega') \mathcal{T}_{\omega',\omega}A_{\Omega,\Phi}(\omega),
    \end{equation}
    \begin{equation}
         \mathcal{R}_{\Upsilon';\Upsilon} = \int_{\mathbb{R}^2_+} d\omega'd\omega\, A^*_{\Omega',\Phi'}(\omega') \mathcal{R}_{\omega',\omega}A_{\Omega,\Phi}(\omega),
    \end{equation}
\end{subequations}
where we use the shorthand notation $\Upsilon=\{\Omega,\Phi\}$, and we omitted the $\sigma$ dependence, which we consider to be of the same value for incoming and outgoing states. If we start from Eq.~(\ref{eq:wdw-full-sol}) integrated over $\int_{\mathbb{R}_+} d\omega A_{\Omega,\Phi} (\omega;\sigma)$, we repeat the steps in the previous section, and we make use of the identity relation $\exp{-\sigma^2/4}/(2\pi)\int d\Omega d\Phi\ A^*_{\Omega,\Phi}(\omega';\sigma)A_{\Omega,\Phi}(\omega;\sigma)=\delta(\omega'-\omega)$, we find
\begin{subequations}
    \begin{equation}
        \mathcal{R}_{\Upsilon';\Upsilon} = iU_{\Upsilon';\Upsilon} + i \int d^2\Upsilon'' U_{\Upsilon';\Upsilon''} \mathcal{R}_{\Upsilon'';\Upsilon}\,,
        \end{equation}
    \begin{equation}
        \mathcal{T}_{\Upsilon';\Upsilon} = \mathcal{I}_{\Upsilon';\Upsilon} + \mathcal{R}_{\Upsilon';\Upsilon}\,,
    \end{equation}
\end{subequations}
where $\mathcal{I}_{\Upsilon';\Upsilon}=\int_{\mathbb{R}_+}d\omega\, A^*_{\Omega',\Phi'}(\omega)A_{\Omega,\Phi}(\omega)$ is the no-interaction term and $d^2 \Upsilon''=(2\pi e^{\sigma^2/4})^{-1} d\Omega'' d\Phi''$. KG current conservation together with wave packets' identity relation shows that $\mathcal{T}$ and $\mathcal{R}$ are well-normalized probability amplitudes w.r.t. the measure $d^2 \Upsilon'$, i.e.,
\begin{equation}
    \int d^2 \Upsilon' (|\mathcal{T}_{\Upsilon';\Upsilon}|^2+|\mathcal{R}_{\Upsilon';\Upsilon}|^2) =1\,.
\end{equation}
Hence, we introduce the following quantities: the probability of a generic FSP and FSR transition
\begin{equation}
    \mathcal{P}^{(\mathcal{\mathcal{M}})} = \int d^2 \Upsilon' |\mathcal{M}_{\Upsilon';\Upsilon}|^2\,,
\end{equation}
with $\mathcal{M}$ denoting either $\mathcal{T}$ or $\mathcal{R}$, and the probability of a FSP/FSR bounce/contraction
\begin{equation}
    \mathcal{P}^{(\mathcal{M})}_{B/C} = \int d^2 \Upsilon'\, \theta(\pm\Phi')|\mathcal{M}_{\Upsilon';\Upsilon}|^2\,,
\end{equation}
where the sign inside the Heaviside theta distinguishes between the bounce (+) and contraction (-). By considering transition probability densities, conditioned on whether the bounce has occurred or not, we can compute $\Omega',\Phi'$ mean values and variances relative to these outcomes 
\begin{subequations}
    \begin{equation}
        \langle \Upsilon'\rangle^{(\mathcal{M})}_{B/C} = \frac{1}{\mathcal{P}^{(\mathcal{M})}_{B/C}} \int d^2 \Upsilon'\, \theta(\pm\Phi')\,\Upsilon' \mathcal{M}_{\Upsilon';\Upsilon}\,,
    \end{equation}
    \begin{equation}
        (\Delta\Upsilon')^{\mathcal{M}}_{B/C} = \sqrt{\langle \left(\Upsilon'\right)^2\rangle^{(\mathcal{M})}_{B/C} - \left(\langle \Upsilon'\rangle^{(\mathcal{M})}_{B/C}\right)^2}\,.
    \end{equation}
\end{subequations}

\bibliography{references}
\end{document}